\documentclass[11pt]{article}
\usepackage{moriond,epsfig}

\def\be{\begin{equation}}
\def\ee{\end{equation}}
\def\bea{\begin{eqnarray}}
\def\eea{\end{eqnarray}}

\newcommand{\jpsi}{\mbox{$J/\psi$}}
\newcommand{\bc}{\mbox{$B_c$}}

\newcommand{\bp}{\mbox{$B^+$}}
\newcommand{\bs}{\mbox{$B_s$}}
\newcommand{\bzero}{\mbox{$B^0$}}
\newcommand{\lambdab}{\mbox{$\Lambda_b$}}
\newcommand{\pt}{\mbox{$p_T$}}
\newcommand{\bcjpsipi} {\mbox{$B_c \rightarrow J/\psi \pi^+$}}

\begin{document}
\vspace*{4cm}
\title{$b$-PHYSICS MEASUREMENTS AT THE TEVATRON: $m$ AND $\Delta m$}

\author{ S. D'AURIA (for the CDF and D0 collaborations)}

\address{Department of Physics and Astronomy, University of Glasgow,\\
University Avenue,
G12 8QQ, Glasgow, U.K.\\
}

\maketitle\abstracts{
We present experimental results on $b$-hadron mass measurements 
and $b$-meson oscillations based on integrated luminosity of 
250 to 450 pb$^{-1}$ of $p\bar{p}$ collisions at $\sqrt{s} = 1.98$~TeV
by the CDF and D0 collaborations at the TeVatron.
The masses of 
$b$-hadrons have been measured precisely by the CDF collaboration 
in decays containing a \jpsi .
A blind search of the decay mode \bcjpsipi\ resulted in a peak of $18.9 \pm 5.4$ 
candidates at a mass value of $6287.0 \pm 4.5 \pm 1.1$ MeV/$c^2$.
A new limit has been set to the decay $B_{d,s}\rightarrow \mu^+\mu^-$.
Both the CDF and D0 collaborations are in the position to put a limit on the 
frequency of the \bs\ oscillations. D0 reports $\Delta m_s > 5.0$~ps$^{-1}$. 
}

\section{Introduction}
This article reviews the recent results on $b$-physics from the experiments D0 and CDF, which 
are presently collecting data from the $p\bar{p}$ collisions produced at the TeVatron 
collider with a 
centre of mass energy of 1.96 TeV. The description of the two experiments and an update on the TeVatron performance
are reported elsewhere 
in these proceedings~\cite{bernardi}. In this paper we review the mass measurements of $b$-hadrons, 
the search  for their rare decays and the search for the \bs\ oscillations.

 A precise measurement of the mass of $b$-hadrons allows testing of the methods used in Lattice QCD and in
potential models. At present all but one ground state of $b$-mesons foreseen by the quark theory 
have been detected and 
for all but two the mass has been experimentally measured with uncertainties below the 
theoretical uncertainty. The mass of the \bc\ meson was calculated using 
non-relativistic potential models~\cite{Eichten:1994gt,Godfrey:2004ya}, Lattice QCD~\cite{Allison:2004be} and 
perturbative QCD calculations~\cite{brambilla:2002}
with an error that is about two orders of magnitude below the experimental uncertainty \cite{CDFbc}.
The CDF collaboration reports here their recent and more precise measurements of the the mass of \bzero,\bp, \bs\ and \lambdab .
Also reported here is the first evidence of the \bc\ meson decay in a fully reconstructed mode, thus allowing the 
measurement of the \bc\ mass with a precision  comparable to the mass of other $b$-mesons.

Both collaborations have updated their upper limit on the branching fractions
of \bs\ and \bzero\ to muon pairs, which would indicate the 
presence of new Physics~\cite{sugra,SO10} if detected at a level
above $\approx 5 \times 10^{-9}$. The D0 collaboration reports the first evidence 
of the 
decay $B_s \rightarrow D_{s1}(2536) \mu +${ \em anything}.
The mass and production yield of excited $b$-mesons have also been observed by both experiments.

 The measurement of the $B_s - \bar{B}_s$ oscillation parameters is one of the main physics goals of
both experiments at the TeVatron. This paper will review the methods used and report on 
the recent limits achieved.

\section{$b$ production cross section and trigger}
The production cross section of $b$ quarks at the TeVatron is approximatively 
a factor 1000 times larger than at the $B$-factories. However, the signal due to $b$ physics 
has to be extracted from a background due to other QCD processes that is 1000 times larger 
than the signal. This is accomplished using three types of triggers: a trigger based on 
muon pairs from the decay of a \jpsi , a trigger based on the detection of a ``soft lepton''
and a trigger based on tracks that are originating from a secondary decay vertex \cite{ash}
that is presently implemented only in the CDF experiment.
The data from the \jpsi\ trigger has been used for mass and lifetime measurements,
the data from the semileptonic and the secondary vertex triggers have been used for 
lifetime\cite{lipton}  and mixing measurement.  
Using the data collected with the \jpsi\ trigger the CDF collaboration has measured~\cite{Acosta:2004yw} the 
single $b$-quark production cross section integrated over one unit of rapidity:
\begin{equation}
 \sigma(pp\rightarrow \bar{b} X, |y| < 1) = (29.4 \pm 0.6 (stat.) \pm 6.2 (syst.) ) ~\mu b
\end{equation}

\section{$b$-hadron masses} 
  The $b$-hadron masses have been measured by reconstructing decays containing a 
muon pair from a \jpsi .
 The two experiments have complementary design features: CDF has a better mass resolution,
while D0 has a larger angular acceptance for muons. The spectrometer of the CDF detector
 has been calibrated using the \jpsi\ mass as a reference by correcting for passive 
material effect, to eliminate the \pt\ dependence of the mass. The actual 
value of the magnetic field  has been tuned 
to obtain the world average \jpsi\ mass value~\cite{PDG}. The calibration 
has been checked against the $\Upsilon$ mass. 
The high statistics of $J/\psi\rightarrow \mu^+\mu^-$  decays available  
has allowed  the CDF experiment to reduce the systematic uncertainties 
to sub-MeV values.
Using the following decay modes and a luminosity of 220 pb$^{-1}$ 
CDF has measured the following preliminary values for hadron masses, all in MeV/$c^2$:

\begin{table}[h]
\begin{tabular} { l l }
$B^\pm \rightarrow J/\psi K^\pm$   & $      m(B^+) = (5279.10 \pm 0.41 \pm 0.36) $ \\
$B^0 \rightarrow J/\psi K^{*0} $   &$       m(B^0) = (5279.63 \pm 0.53 \pm 0.33)  $ \\
$B_s \rightarrow J/\psi \phi    $   &  $      m(B_s) = (5366.01 \pm 0.73 \pm 0.33)  $\\
$\Lambda_b \rightarrow J/\psi \Lambda^0 $&$ m(\Lambda_b) = (5619.7 \pm 1.2 \pm 1.2) $\\
\end{tabular}
\label{test}
\end{table}

\section{Evidence of the  decay $B_c^\pm \rightarrow J/\psi \pi^\pm$ and \bc\ mass measurement}
The \bc\ meson has been observed~\cite{CDFbc} at the TeVatron Run I by the CDF collaboration in decays containing a 
neutrino, and therefore 
its mass has a large experimental uncertainty. 
The D0 collaboration has recently observed the semileptonic decays of the \bc\ meson 
in the Run 2 data~\cite{d0bc_ichep}.
The LEP experiments  searched for the fully reconstructed decays of the \bc\ 
meson~\cite{Ackerstaff:1998zf,delphi_bc,Barate:1997kk}, but the \bc\ production cross section
from $Z$ decays was too low for detecting these decays.
The CDF collaboration has now evidence for 
the fully reconstructed decay mode  $B_c^\pm \rightarrow J/\psi \pi^\pm$, with 
$J/\psi\rightarrow \mu\mu$. This simple decay mode has a relatively large expected 
branching ratio~\cite{kiselev} when the daughter decays are also included. 
It is detected with a trigger
that is relatively efficient and is not based on the decay vertex.
The CDF collaboration has used an analysis technique that blinded a wide range of the 
mass distribution during cut optimization. 
The statistical 
test that was used to assess the significance of the signal
was completely set before unblinding the 
mass distribution. The test was based on the score function
$\Sigma = N_s / (1.5 +\sqrt{N_b})$ where $N_s$ and $N_b$ are the number of 
signal and backgorund candidates as obtained from a fit in a sliding mass window that was 
300~MeV/c$^2$ wide.
As the mass value was known with an uncertainty of 
$\pm 400$~MeV/c$^2$ the mass peak 
corresponding to this decay was searched for in the range 
$5700 \le M(B_c) \le 7000$~MeV/c$^2$. 
The analysis required a good fit to a displaced 
vertex using well defined tracks with silicon hits, 
making use of  the innermost
 silicon layer (L00). 
Upon unblinding the mass distribution, it was found that one region 
contained a peak that satisfied the 
predefined statistical test. 
The probability that the background generates a fluctuation equal or larger
than the observed signal, anywhere in the search region, was estimated using 
Monte Carlo generated distributions
that simulated only the background. This probability was found a posteriori 
to be about 0.27\%.
The experimental evidence of this decay mode allowed measuring the \bc\ mass.
The $J/\psi\pi$ mass distribution is shown in fig.~1. The unbinned likelihood fit
returns $18.9\pm 5.4$ \bc\ candidates on a background of $10.1\pm 1.4$ events.
The main contribution to the systematic uncertainties comes from fitting
with different background shapes. Other systematic uncertainties are derived from the
mass measurements reported above. The experimental value of the \bc\ mass is
\begin{equation}
m(B_c) = 6287.4 \pm 4.5(stat.) \pm 1.1 (syst.) \mbox{~MeV/c}^2
\end{equation}
The details of this analysis will be published in a forthcoming paper~\cite{bcprl}.
This result is in very good agreement with the theoretical expectations mentioned above
~\cite{Eichten:1994gt,Godfrey:2004ya,Allison:2004be,brambilla:2002}.
\begin{figure}[htbp]
  \begin{center}
    \includegraphics[width=12.0cm]{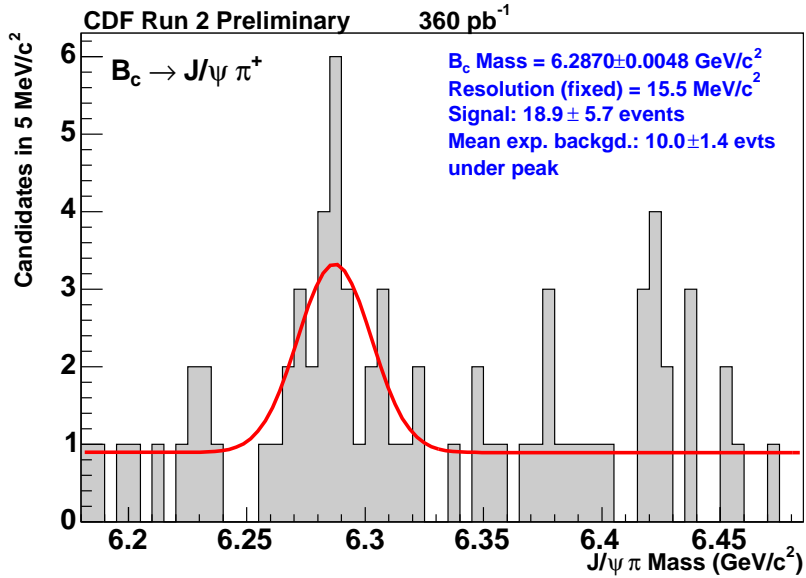}
    \caption{Mass distribution of the \bc\ candidates. The fit is a gaussian function with a linear background.}
  \end{center}
\label{fig:bcmass}
\end{figure}
Checks on the detection of the partially reconstructed decays $B_c \rightarrow J/\psi~+$~{\em track} 
$+$ {\em anything} 
in the same sample also gave a positive result, with a significant excess only for $m(J/\psi\pi) < m(B_c)$, in agreement 
with the expectations.
\begin{figure}[htbp]
  \begin{center}
    \includegraphics[width=12.0cm]{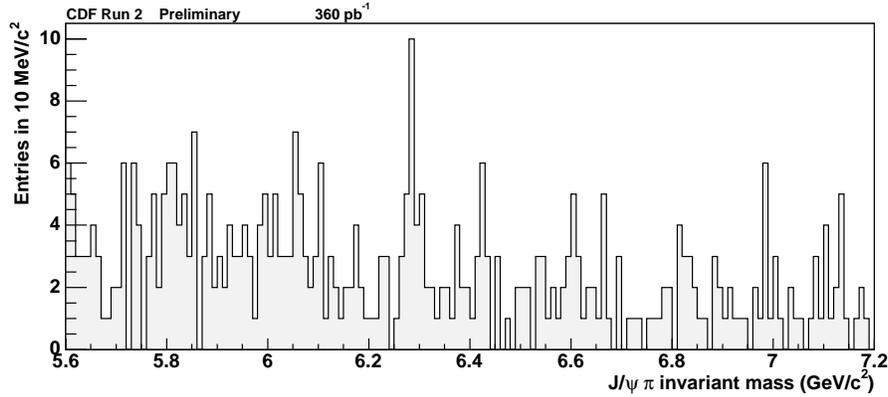}
    \caption{
  Mass spectrum for the \jpsi\ $\pi^\pm$ candidates in the whole search region for the \bcjpsipi .
}
  \end{center}
\label{fig:bcfullrange}
\end{figure}

\section{Excited B mesons}
The excited states of the $b$-mesons are of great interest to measure their mass, branching ratios and
width, which are predicted with a good precision by various theoretical models.
In addition,
at the hadron collider the excited states of the $b$ mesons must be taken into account in mixing measurements,
when using a flavour tagging method that is based on the information from tracks that are near to the candidate $b$ meson.
The excited B states had been studied at LEP by the Aleph collaboration~\cite{Barate:1998cq}.
The D0 collaboration has measured the mass of the two narrow states corresponding to the orbital angular momentum 
L=1 $B_1$ and $B_2^*$ in their decay to $B^{(*)} \pi$. 
Also the CDF collaboration confirms evidence of two separate peaks. In both experiments fully reconstructed decays of the
\bp\ and  \bzero\ mesons are detected and are associated with a track from the primary vertex. 
 The three contributions are from the decays $B_1\rightarrow B^* \pi$,
$B_2^*\rightarrow B^* \pi$ and $B_2^*\rightarrow B \pi$, with $B^*\rightarrow B\gamma$. The undetected photon of about 46 MeV
causes the larger shift in the mass peak. The D0 preliminary results are~\cite{d0bc_ichep}:   
\begin{equation}
m(B_1) = 5274 \pm 4(stat.) \pm 7(syst.) MeV/c^2
\end{equation}
\begin{equation}
m(B_2^*) - m(B_1) = 23.6 \pm 7.7(stat) \pm 3.9 (syst) MeV/c^2
\end{equation}
\begin{equation}
\Gamma_1=\Gamma_2 = 23 \pm 12(stat) \pm 9 (syst) MeV
\end{equation}

\section{Search for rare decays}
The decay of a neutral $b$-meson into two opposite charged muons
is mediated in the standard model by 
flavour changing neutral currents due to ladder diagrams and the corresponding
branching ratio is calculated to be~\cite{Misiak:1999yg,Buchalla:1998ba}
 $BR(B_{d,s}\rightarrow \mu^+\mu^-) = (3.4\pm0.54) \times 10^{-9}$. 
A branching ratio that is larger than this value would indicate the presence of
new Physics processes, that can considerably enhance  this decay channel. The D0 collaboration has
searched for the decay of $b$-mesons to two muons using data from an integrated luminosity of 
300~pb$^{-1}$. Four  candidates were found in the mass window, giving a limit on this branching ratio 
$BR(B_{d,s}\rightarrow \mu^+\mu^-) \leq 3.7 \times 10^{-7}$ at 95\% conficence level. 
\begin{figure}[htbp]
  \begin{center}
    \includegraphics[width=12.0cm]{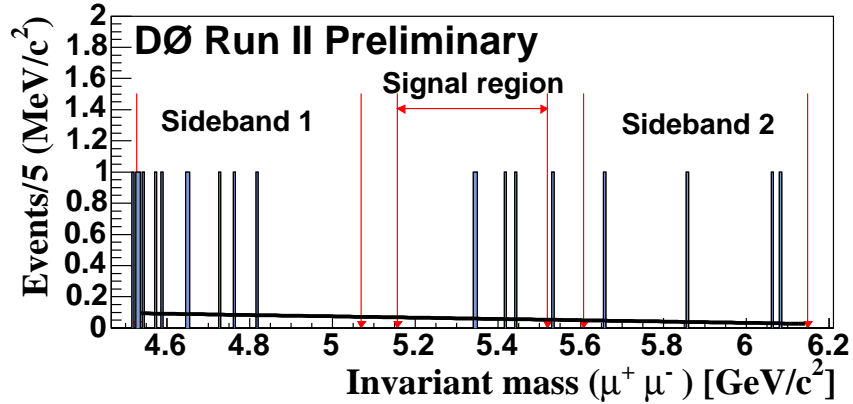}
    \caption{
 Invariant mass distribution of the muon pairs. Four candidates are found in the signal region.
}
  \end{center}
\label{fig:btomumu}
\end{figure}
The CDF collaboration uses its better 
detector resolution to put separate limits on the decay rate of the two mesons~\cite{Acosta:2004xj}: 
$BR(B_{s}\rightarrow \mu^+\mu^-) \leq  5.8 \times 10^{-7}$ and $BR(B_{d}\rightarrow \mu^+\mu^-) \leq  1.5 \times 10^{-7}$ at 90\% 
confidence level using a collected luminosity of 171~pb$^{-1}$. This result has been updated
shortly after this conference.
The D0 collaboration has also measured the sensitivity  to detect a signal from the decay
$B_s\rightarrow\mu\mu\phi$. Using data from 
300~pb$^{-1}$ they can put a limit on the branching fraction  $ BR \leq 1.2 \times 10^{-5}$ at 95\% CL.

\section{Flavour oscillation measurements}
To perform time-dependent mixing measurements at the hadron collider 
it is necessary to select a 
flavour--specific decay signal (e.g. $B_s \rightarrow D_s \pi^-$),
establish the flavour of the $b$-meson at production and measure precisely the proper decay time, 
which involves 
measuring the decay length and the momentum of the reconstructed meson.
The relevant parameters are the purity of the sample ($f_{sig} = N_{signal}/N_{background}$), which indicates 
the fraction of decay signal in the sample,
the tagging efficiency $\epsilon$, which indicates the fraction of the signal sample that 
has a flavour tag, $\epsilon = \frac{N_{tag}}{N_{cand}}$, and the dilution $D$,
which indicates the probability that the tag is correct: $D = \frac{N_R - N_W}{N_R + N_W}$
The statistical power of the sample is diluted by a factor $\epsilon D^2$: the significance
of the oscillation signal is given by~\cite{PDG}: 
\begin{equation}
S \propto f_{sig} \sqrt{1/2 ~\epsilon D^2 N_{cand}}~  
 ~e^{- \frac{1}{2} ( \Delta m_s \sigma_t)^2} 
\end{equation}
where $\Delta_m$ is the mixing parameter that determines the frequency of the oscillations
and $\sigma_t$ is the proper time resolution.  The latter is the sum in quadrature of two terms:
one related to the spatial resolution of the secondary vertex 
and the second related to the momentum resolution of the reconstructed meson.

For $B_s$ mixing we have a lower statistics compared to the \bzero\ channel and a larger
oscillation parameter, which implies faster oscillations. Therefore, to have a large significance 
for large $\Delta m_s$ we need a very good proper time resolution, which can be better achieved with
fully reconstructed decays, at the price of smaller statistics.

Both collaborations have used, until now, only the ``opposite side'' tagging methods, which rely
on tagging the flavour of a $b$-meson by looking at the characteristics of particles
that are produced ``away'' from the reconstructed meson and that presumably contain hadrons from
the $b$-quark that was produced in association with the one that originated the reconstructed 
$b$-meson. 
This tagging method is independent on the nature of the meson 
under study, so it has been tested on the \bzero\ mixing and applied to the search for \bs\ mixing.

\section{\bzero\ Mixing}
The D0 collaboration has used only the semileptonic decays $B^0\rightarrow D^{*\pm}\mu^\mp  X$,
(Here $X$ denotes any  set of particles produced in the decay). 
The CDF collaboration has analysed some fully reconstructed decays from the vertex trigger, 
for which  the lifetime bias is now well understood~\cite{lipton}. Both collaborations have results that 
are in very good agreement with the precise measurements, which have been performed at the $B$-factories,
as shown in Table~1. The purpose of these measurement was mainly to test the fit algorithms and to 
obtain the dilution factors, to be used in the search for \bs\ oscillations.

\begin{table}[t]
\caption{
Values of $\Delta m_d$ obtained by CDF and D0 using only the opposite-side flavour tag, 
for semi-leptonic decays (S.L.) and
hadronic decays (Had).
\label{tab:bzeromix}
}
\vspace{0.4cm}
\begin{center}
\begin{tabular}{|l|l|l|}
\hline
& &  \\
& $\Delta m_d$ (ps$^{-1}$) & $\epsilon D^2$ (\%)\\ \hline & &  \\ 
D0 (S.L.) & $0.558 \pm 0.048$ (stat.) & $ 1.6 \pm 0.05$ (stat.)   \\ \hline & &  \\
CDF (S.L.) & $0.497 \pm 0.028$ (stat.)$\pm 0.015$ (syst.) &$  1.43 \pm 0.09$ (stat.+syst.) \\ \hline & &  \\
CDF ({\em Had.})& $0.503 \pm 0.063 $(stat.)$\pm 0.015$(syst.) & $1.12\pm 0.18$(stat.) $\pm0.04$(syst.)\\
& &   \\ \hline
\end{tabular}
\end{center}
\end{table}


\section{\bs\ Mixing}
The D0 collaboration has applied an updated version of 
the opposite side muon tagging algorithm to the signal sample enriched in \bs\ 
semileptonic decays $B_s \rightarrow D_s \mu X$, with $D_s^\pm \rightarrow \phi \pi^\pm$. 
Using $376\pm31$ reconstructed and tagged decays (on a total of 7037), 
and with a tagging dilution of
$D =  0.552 \pm 0.016$ a very
 preliminary fit gives a
null result on oscillations, with a limit to the parameter 
$\Delta m_s \geq 5.0$~ps$^{-1}$ at 95\% confidence level 
and a sensitivity of 4.6~ps$^{-1}$, using both 
the statistic and the systematic uncertainties.
This result is not as good as the world average, that indicates $\Delta m_s \geq 14.5$~ps$^{-1}$
but it is only the first attempt to fit the data. A considerable improvement
is expected, both on statistics and on tagging efficiency. In particular, using the
same-side-tagging techniques the tagged sample can be increased using the same data
collected to date. 
\begin{figure}[htbp]
  \begin{center}
    \includegraphics[width=12.0cm]{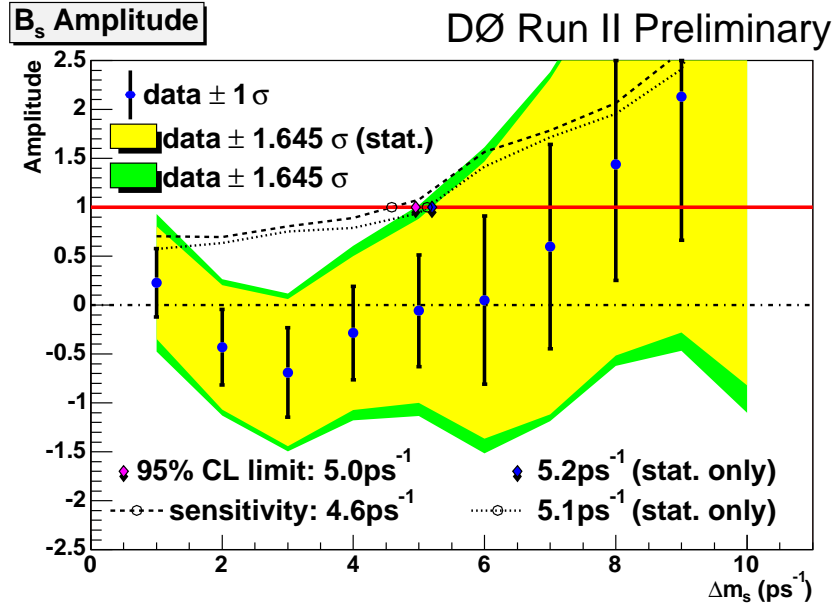}
    \caption{
 Amplitude scan analysis and  $\Delta m_s$ limit from the D0 collaboration.
}
  \end{center}
\label{fig:bsmixing}
\end{figure}

By comparison, the CDF collaboration has obtained similar results in terms of $\epsilon D^2$. 
 The CDF sample has about 7500 candidates to \bs\ semileptonic decays, in three reconstructed decay 
modes of the
$D_s$: $\phi \pi^+$, $K^{*0} K^+$ and $\pi^+\pi^-\pi^+$.
For fully--reconstructed hadronic decay candidates $B_s \rightarrow D_s \pi^-$, with 
the $D_s$ decaying to the same modes as above, the statistics is
$\approx 900$ candidates; the expected tagging power $\epsilon D^2$ factors are reported in Table~1.

\section{Other results}
The D0 collaboration has found evidence for the decay
$B_s \rightarrow D_{s1}(2536) \mu +X$, heading to a measurement of this
 branching ratio and of the mass of the $ D_{s1}(2536)$.

\section{Conclusions and prospectives}
The experiment at the Tevatron have considerably 
improved the precision on $b$-mesons mass measurements. In particular, the \bc\ 
mass has been measured with good precision thanks to the detection of a signal in a fully
reconstructed mode, which is reported here for the first time. 
Also the masses of two B-mesons excited states has been measured.
The mixing parameter $\Delta m_d$ is in good agreement with the world average, 
but an initial look at the \bs\ oscillations has given a null result on the measurement of 
$\Delta m_s$, with a limit that is still more than a factor of two lower than the world average.
Large improvements are expected not only with more statistics, but also with
improved techniques in tagging, improving vertex resolution and adding other 
decay channels.

\end{document}